\documentclass[journal=jacsat,manuscript=article]{achemso}
\usepackage[version=3]{mhchem} % Formula subscripts using \ce{}
\usepackage{amssymb}
\usepackage{graphicx}

\newcommand{ \be}{\begin{equation}}
\newcommand{ \ee}{\end{equation}}
\newcommand{\beq}{\begin{eqnarray}}
\newcommand{\eeq}{\end{eqnarray}}
\newcommand{\bem}{\begin{pmatrix}}
\newcommand{\eem}{\end{pmatrix}}
\newcommand{\bmx}{\begin{array}}
\newcommand{\emx}{\end{array}}

\author{Boris A. Klumov}
\affiliation{Joint Institute for High Temperatures, Moscow, 125412, Russia}
\email{klumov@ihed.ras.ru}
\affiliation{Moscow Institute of Physics and Technology, Moscow, Russia}

\author{Yuliang Jin}
\affiliation{Levich Institute, Physics Department, New York, NY, 10031, USA}

\author{Hernan A. Makse}
\affiliation{Levich Institute, Physics Department, New York, NY, 10031, USA}

\title{Structural properties of dense hard spheres near random close packing}

\abbreviations{IR,NMR,UV}

\begin{document}

\begin{abstract}
We numerically study structural properties of mechanically stable packings of hard spheres (HS), in a wide range of packing fractions $0.53 \le \phi \le 0.72$. Detailed structural information is obtained  from the analysis of orientational order parameters, which clearly reveals a disorder-order phase transition at the random close packing (RCP) density, $\phi_{\rm c} \simeq  0.64$.  Above $\phi_{\rm c}$ the crystalline nuclei form 3D-like clusters, which upon further desification, transform into alternating
planar-like layers. We also find that particles with icosahedral symmetry survive only in a narrow density range in the vicinity of the RCP transition.
\end{abstract}

\section{Introduction}

As a fundamental model in condensed matter physics and material science, packings of hard spheres (HS) reproduce many essential structural properties of glassy and granular  media \cite{rmp1,rmp2,ncom}. Structural changes of the HS system have been observed when a packing is densified above the density of random close packing (RCP) $\phi_{\rm c}  \simeq 0.64$ (see, e.g. \cite{hs_prb, makse}).  A new order parameter, which is based on the cumulative distribution of the rotational invariant $w_6$, was proposed  in \cite{hs_prb} to identify this structural changes. Random arrangement is transferred to partial crystallization across RCP, and two lattice types, the face centered cubic (fcc) and the hexagonal close-packed (hcp), are observed in the crystalline clusters. However, more detailed information has not been obtained due to insufficient system size. In this study, we aim to explore more detailed structural changes of the transition at RCP in comparison with previous studies \cite{hs_prb, makse}, based on larger scale simulations.

We numerically generate a large set of HS packings composed of $N=64 \times 10^3$ monodispersed spheres 
located in the cubic box with periodic boundary conditions, using the modified  Lubachevsky-Stillinger (LS) algorithm \cite{jtb}.
%The packing fraction depends on the compression rate as slower compressions result in denser packings. The packings obtained from LS algorithm are then used as initial configurations to generate mechanically stable jammed packings of particles interacting with Hertz normal forces. When $\phi < 0.64$, Mindlin frictional forces are added to stabilize the packings. By changing the compression rate in the LS algorithm, we obtain packings with a wide range of packing fractions, $\phi \simeq 0.53 - 0.72$.  The range includes the random close packing state at $\phi_{\rm c}\simeq 0.64$ (Bernal limit) \cite{bernal}.  More details of the numerical algorithm are described in Ref.~\cite{makse}.

\subsection{Results}
\begin{figure}
\includegraphics[width=12.cm]{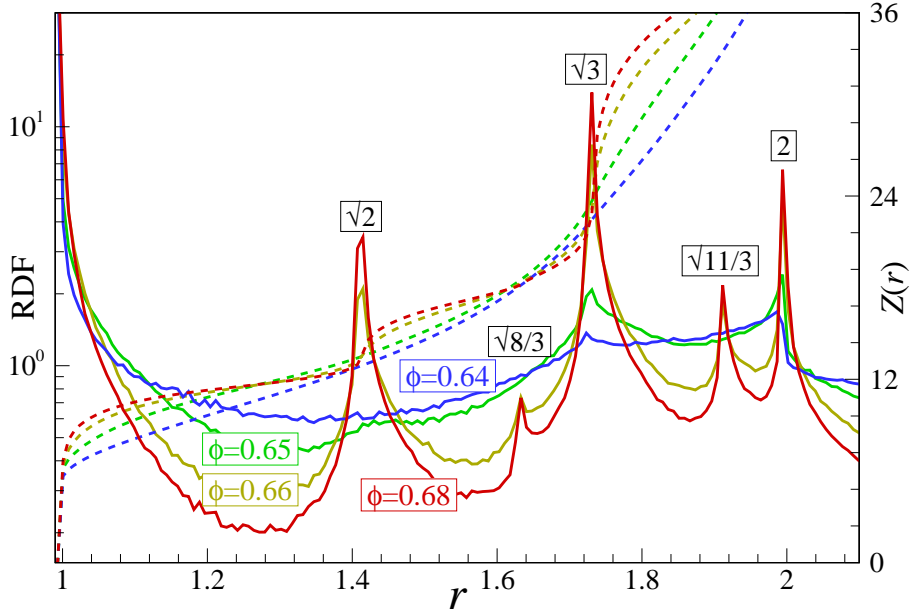}
\caption{(Color online) Radial distribution function $g(r)$ of hard spheres (solid lines) and its cumulative function $Z(r)$ (dashed lines) at different packing fractions $\phi$. The values  of $\phi$, and the positions of peaks, are indicated on the plot.}
%{\bf Boris, please indicate in the plot the peak at $\sqrt{11/3}$}}
%Grey line corresponds to $Z(r) = 12$.}
\label{rdf}
\end{figure}

To examine the structural properties of jammed spheres, we first study their spatial correlations. Figure~\ref{rdf} shows the radial distribution function $g(r)$ of hard spheres for several different packing fractions in the range $[0.60, 0.68]$. The cumulative function $Z(r) \equiv 4\pi \rho \int_0^r r'^2 g(r') dr'$, which is the mean number of particles inside a sphere of radius $r$, is also plotted in Fig.~\ref{rdf}. When $\phi \le \phi_{\rm c}$, addition to the strong peaks at the contact distances $r=1,2,\dots$ (we set unit diameter), a weak secondary peak at $r = \sqrt{3}$ is visible. This peak corresponds to an contact angle $\theta = 2\pi/3$, but not necessarily a lattice structure ~\cite{Donev2005}. Above $\phi_{\rm c}$, additional peaks appear at $\sqrt{2}$, $\sqrt{8/3}$, and $\sqrt{11/3}$, which are typical distances in the fcc and hcp lattices. The observation indicates the onset of crystallization when the packings are densified above $\phi_{\rm c}$.

More detailed structural information can be obtained from well known bond order parameter method \cite{stein}, which is widely used to study the structural properties of condensed matter\cite{stein,torqa,tw}, hard spheres\cite{hs_init,troadec,iv,aste,anik,makse,hs_prb,kapf,franc,bar}, Lennard-Jones systems \cite{q6a,torqb,lj1,lj2}, complex plasmas\cite{nat,3Da,mitic,3Db,pu,khr}, colloidal and patchy systems\cite{coll0,coll1,pc}, granular  media\cite{gran}, metallic glasses\cite{mg}, repulsive shoulder systems\cite{rss}, etc.
Each particle $i$ is connected via vectors (bonds) with its $N_{\rm nn}(i)$ nearest neighbors (NN), and the rotational invariants (RIs) of rank $l$ of second $q_l(i)$ and third $w_l(i)$ orders are calculated as:
\be
q_l(i) = \left ( {4 \pi \over (2l+1)} \sum_{m=-l}^{m=l} \vert~q_{lm}(i)\vert^{2}\right )^{1/2}
\ee
\be
w_l(i) = \hspace{-0.8cm} \sum\limits_{\bmx {cc} _{m_1,m_2,m_3} \\_{ m_1+m_2+m_3=0} \emx} \hspace{-0.8cm} \left [ \bmx {ccc} l&l&l \\
m_1&m_2&m_3 \emx \right] q_{lm_1}(i) q_{lm_2}(i) q_{lm_3}(i),
\label{wig}
\ee
\noindent
where $q_{lm}(i) = N_{\rm nn}(i)^{-1} \sum_{j=1}^{N_{\rm nn}(i)} Y_{lm}({\bf r}_{ij} )$, $Y_{lm}$ are the spherical harmonics and ${\bf r}_{ij} = {\bf r}_i - {\bf r}_j$ are vectors connecting centers of particles $i$ and $j$. In Eq.(\ref{wig}), $\left [ \bmx {ccc} l&l&l \\ m_1&m_2&m_3 \emx \right ]$ are the Wigner 3$j$-symbols, and the summations performed over all the indexes $m_i =-l,...,l$ satisfying the condition $m_1+m_2+m_3=0$. As shown in the pioneer paper\cite{stein}, the bond order parameters $q_l$ and $w_l$ can be used as measures, to characterize the local orientational order and the phase state of considered systems. Because each lattice type has a unique set of bond order parameters, the method of RIs can be also used to identify lattice structures in mixed systems. The values of $q_l$ and $w_l$ for a few common lattice types (including liquid-like HS state) are presented in Table~\ref{t1}.
Some of the values were reported previously see, e.g. Refs.~\cite{stein,t1a,iv,t1b,t1c,t1d,3Da}. From Eq.~(\ref{wig}), it is easy to see that $w_l \propto q_l^3 $, and therefore we expect $w_l$'s to be more sensitive measures compared to $q_l$'s.

\begin{table}[!ht]
\centering
\caption{Bond order parameters $q_l$ and $w_l$ ($l=4,~6$) for several typical lattices and liquid-like HS random packings (rp) at $\phi=0.55$, calculated via the fixed number of nearest neighbors (NN).}
\begin{tabular}{|c|c|c|c|c|}
%\begin{tabular}{ll|cccc}
\hline %\hline
lattice type & \quad $q_{4}$ & \quad $q_{6}$ & \quad $w_{4}$ & \quad $ w_{6}$
\\ \hline hcp (12 NN) & 0.097 & 0.485 & 0.134  & -0.012
\\ \hline fcc  (12 NN) & 0.19  & 0.575  & -0.159 &  -0.013
\\ \hline ico  (12 NN) & $1.4 \times 10^{-4}$ & 0.663 & -0.159  & -0.169
\\ \hline bcc ( 8 NN) & 0.5 & 0.628 & -0.159   & 0.013
\\ \hline bcc (14 NN) & 0.036 & 0.51 & 0.159   & 0.013
\\ \hline rp at $\phi =0.55$ (12 NN) & $\approx $0.16  & $\approx $0.34  & $\approx $-0.019  &$\approx $-0.032
\\\hline %\hline
\end{tabular}
\label{t1}
\end{table}

\begin{figure}
\includegraphics[width=10.cm]{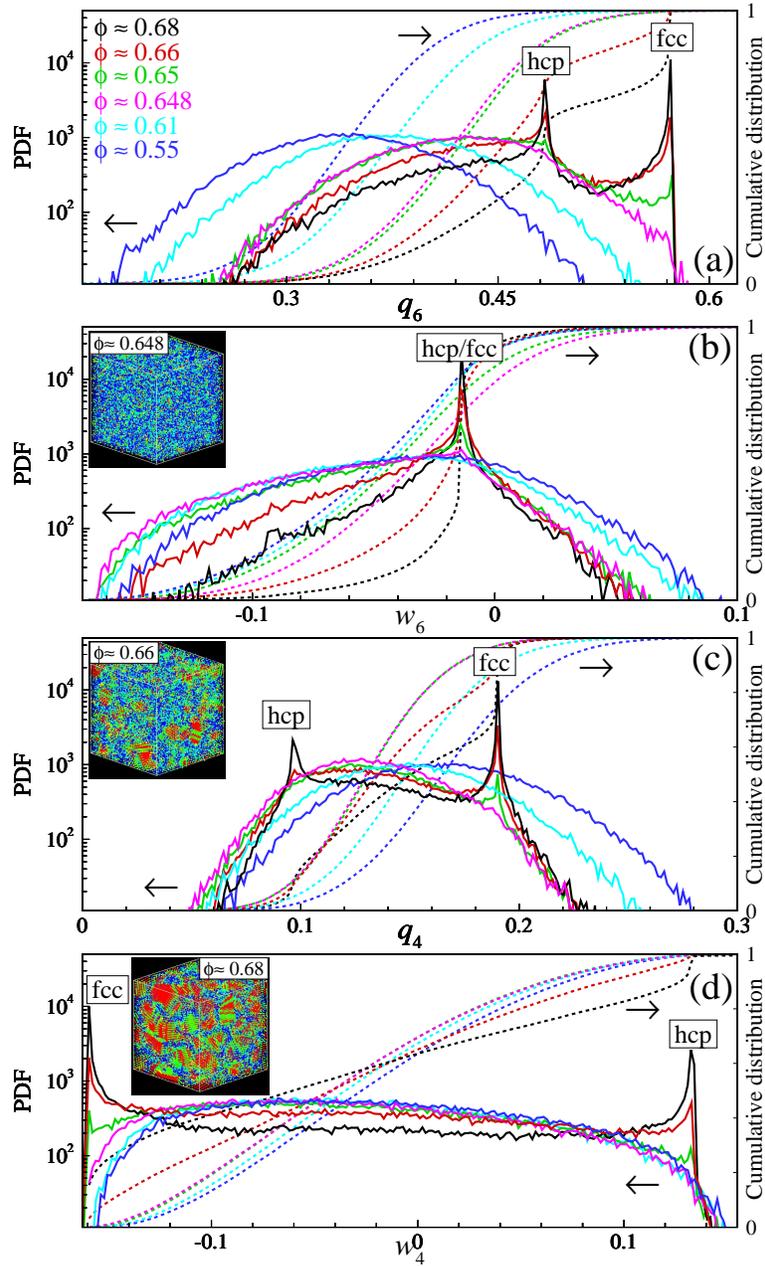}
\caption{(Color online) Probability distribution functions (PDFs) of the rotational invariants $q_i$ (a and c) and $w_i$ (b and d) at various packing fractions $\phi$. Values of the bond order parameters for ideal lattice types (fcc and hcp) are also indicated. Cumulative distributions (normalized to unity) of these PDFs are shown by dashed lines, which are much less noisy than PDFs. By using these distributions  it is easily to estimate the abundance of crystalline and liquid-like particles. Insets (from top to bottom) show spatial distributions of $N = 64\times 10^3$ HS in the cubic box at different packing fractions $\phi \simeq 0.65,~ 0.66,~0.68$, respectively. Hard spheres are color-coded by their $q_6$ values, in order to visualize liquid-like (blue color) and crystalline hcp-like (green color) and fcc-like (red color) structures.}
\label{pdf}
\end{figure}

In three dimensions (3D) the densest possible packings of identical hard spheres are the fcc and the hcp lattices (with the densest packing fraction (pf) $\phi_{\rm fcc} = \phi_{\rm hcp} = \sqrt{2}\pi/6 \simeq 0.74$).  Dense HS systems may also include particles having the icosahedral (ico)  type of symmetry. Icosahedral, fcc and hcp spheres have 12 nearest neighbors (the spheres located in the first coordination shell).

To identify  lattice-like particles, we calculate the bond order parameters  $q_l$ and $w_l$ ($l=4,~6$) for each particle with a fixed number of the nearest neighbors $N_{\rm nn}=12$ \cite{neighbor}. A particle is called as fcc-like (hcp-like, ico-like) if its coordinates in the 4-dimensional space $(q_4,q_6,w_4,w_6)$ are sufficiently close to those of the perfect fcc (hcp, ico) lattice type. An amorphous (liquid-like) particle is identified if its order parameters are sufficiently small, for example,  $q_6^{\rm liq} \simeq N_{\rm nn}^{-1/2} \simeq 0.29 \ll q_6^{\rm fcc/hcp/ico}$.  Note that by varying NN number and rank $l$ of parameters $w_l$ and $q_l$, in principle it is possible to find any lattice type. For instance, the first and the second (next nearest neighbors) shells of body centered cubic (bcc) lattices correspond to $N_{\rm nn} = 8$ and 14 respectively.

\begin{figure}
\includegraphics[width=12.cm]{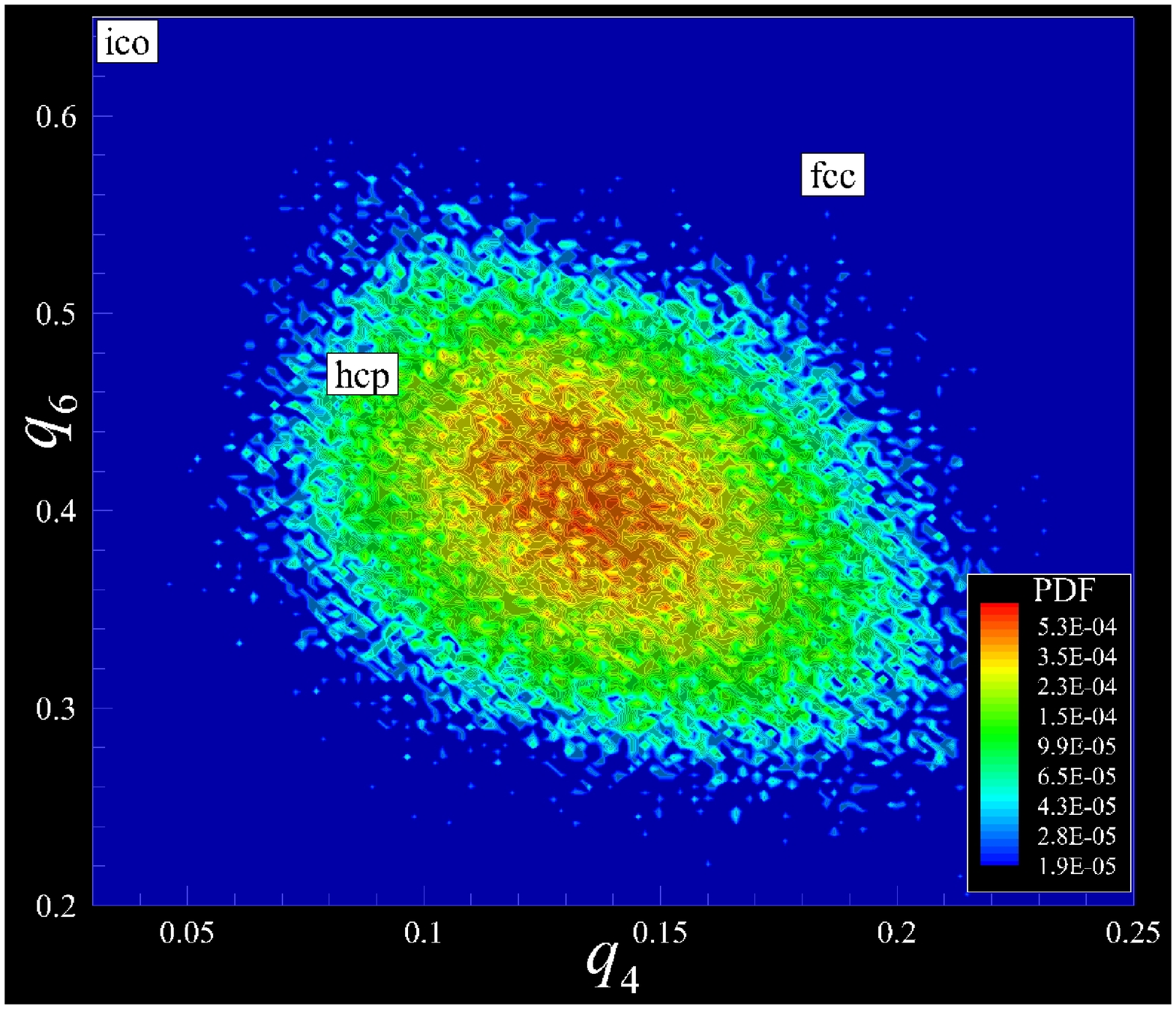}
\includegraphics[width=12.cm]{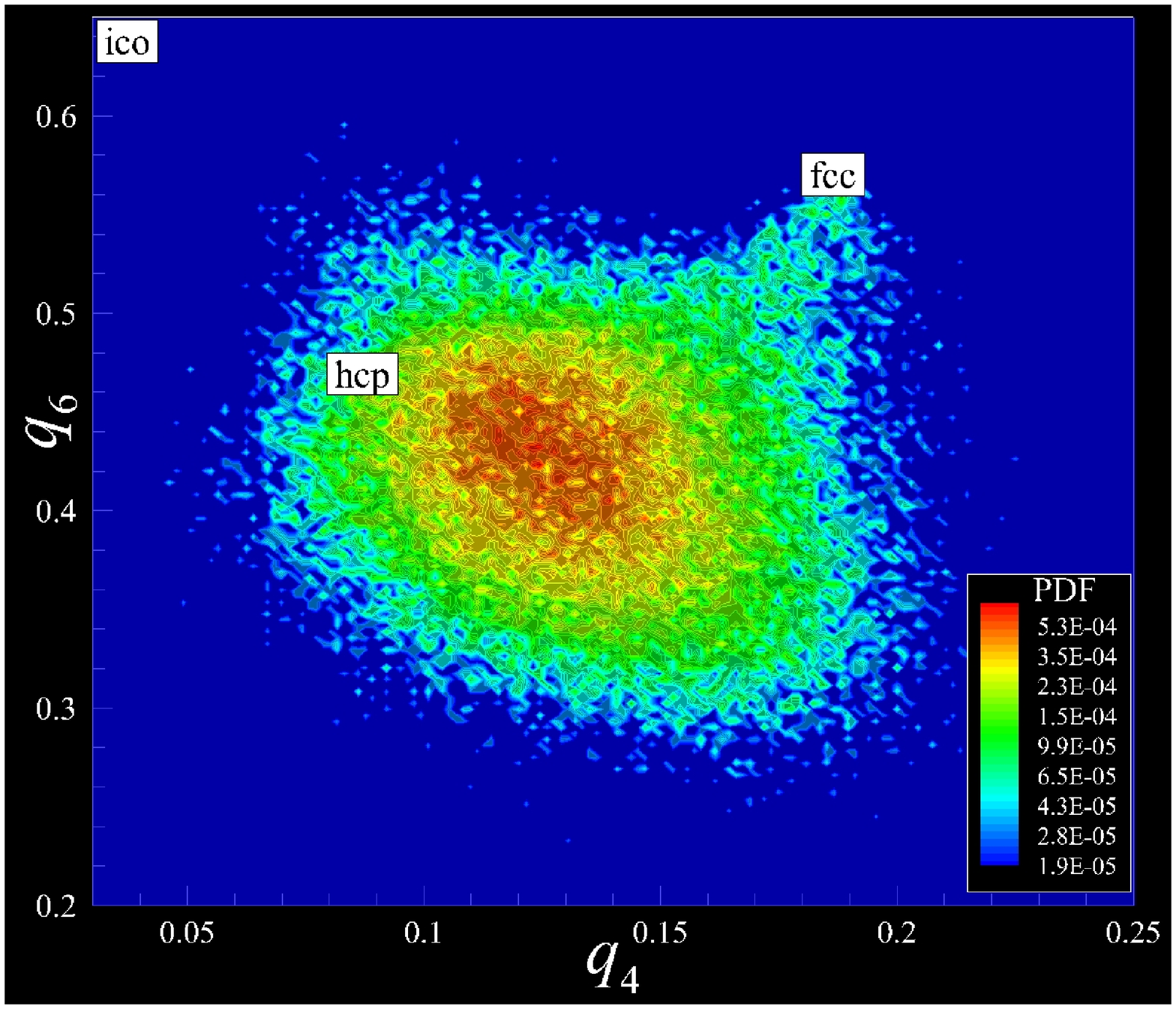}
\caption{(Color online). Probability distributions on the order parameter plane of $q_4 \-- q_6$  in the vicinity of the RCP at $\phi = 0.64$ (top) and $\phi = 0.65$ (bottom). Loci of the perfect fcc, hcp and ico are also indicated on the plot. A fcc peak appears at $\phi = 0.65$. Note that the hcp lattice may have large overlapping with the liquid-like structure in this plot.}
\label{pdf2d}
\end{figure}

%The cumulative function$Z(r)$ shows a mean number of particles inside a sphere of radius $r$, so the Fig~\ref{rdf} clearly reveals that the first coordination shell includes 12 NN in the considered range of $\phi$ values of HS.

To quantify the local orientational order, it is convenient to use the probability distribution functions (PDFs) $P(q_l)$ and $P(w_l)$. Figure~\ref{pdf} shows how the PDFs vary at different $l$ ($l=4,~6$) and $\phi$ (covering both amorphous and solid-like states).  Densification of the HS, as clearly seen in the Figure~\ref{pdf} results in appearance of the crystalline fcc-like and hcp-like spheres in the vicinity of the RCP.  Corresponding cumulative distributions  $C_q^l$ and $C_w^l$ (shown by dashed lines of the same color) are also plotted in the Figure~\ref{pdf}. For instance, the cumulative function $C_q^l$ associated with the $P(q_l)$ is defined as:
\be
C_q^l (x) \equiv \int_{-\infty}^x P(q_l)dq_l
\ee
Evidently, $C_q^l (x)$ is the abundance of particles, having values of $q_l < x$ and $C_q^l (\infty)
= 1$.
Figure~\ref{pdf2d} shows two dimensional PDFs - probability distributions on the plane $q_4 \-- q_6$ - in the vicinity of the RCP at $\phi \simeq 0.64$ and $\phi \simeq 0.65$. At $\phi \simeq 0.64$ structurally HS is amorphous, while at $\phi \simeq 0.65$ the formation of crystalline fcc-like spheres is clearly seen. The result suggests a disordered-ordered transition in the density range [0.64, 0.65],  consistent with previous studies\cite{anik, makse,hs_prb} .

\begin{figure}
\includegraphics[width=14.cm]{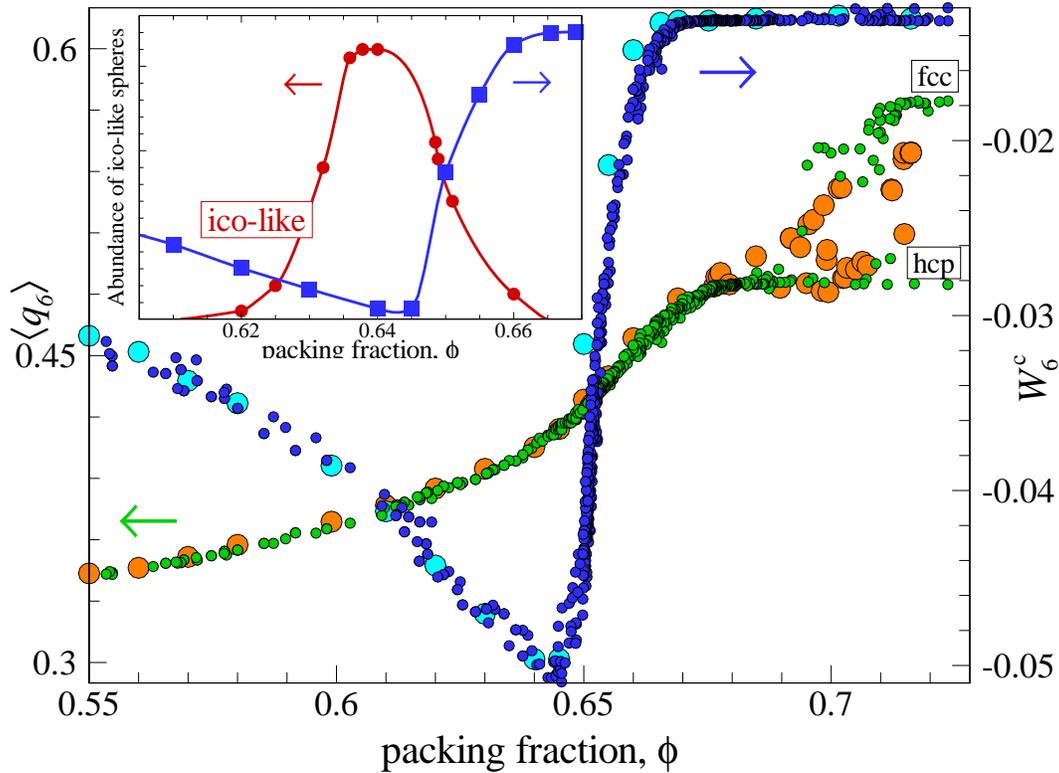}
\caption{(Color online) Cumulant $W_6^c$ (blue and cyan circles) and $\langle q_6 \rangle$ (green and orange circles) versus packing fraction $\phi$. Blue and green circles correspond to HS system with $N=10^4$ spheres; cyan and orange circles correspond to $N=64 \times 10^3$ spheres.  The minimum of $W_6^c$ can be used to locate the RCP transition, $\phi_{\rm c}\simeq 0.645$. The explosive-like growth of $W_6^c$ above $\phi_{\rm c}$ is a clear indicator of the appearance of ordered, crystalline-like (hcp-like and/or fcc-like spheres). The growth of $\langle q_6 \rangle$ value near $\phi_{\rm c}$ is rather monotonous and relatively slow. The possibility for the dense HS system to be packed  via different abundance of fcc-like and hcp-like particles is the key reason behind the fluctuation of $\langle q_6 \rangle$ when $\phi \gtrsim 0.68$. Inset shows the number of ico-like spheres versus $\phi$: ico-like spheres survive only in the narrow range of parameter $\phi$ near RCP transition. }
\label{op}
\end{figure}

The global order parameters, ie., the cumulant $W_6^c$  (defined as $C_w^l (W_l^c) = 1/2$) and the mean value of $q_6$ (defined from the global averaging $\langle q_6 \rangle = (1/N)\sum\limits_{i=1}\limits^N q_6(i)$), are shown in Figure~\ref{op}.  No significant finite size effects are observed  between $N=10^4$ and $N=64 \times 10^3$ systems. For the purpose of detecting crystallization onset, we find $W_6^c$ to be a more sensitive measure than $\langle q_6 \rangle$.  This is because (1) liquid-like and lattice-like spheres have quite different values of $w_6$, and (2) $w_6$ is nearly the same for perfect fcc and hcp lattices (see Table~\ref{t1}). 
Because of this reason, the measure $W_6^c$ has been recently used to describe the melting and freezing phase transition of Lennard-Jones\cite{lj1,lj2} and Yukawa systems\cite{pu,khr,khr1}. On the other side, the density dependence of $\langle q_6 \rangle$ also reveals fascinating properties. The relatively slow increase of $\langle q_6 \rangle$ at $\phi \lesssim \phi_{\rm c}$ can be attributed to single hcp-like particles\cite{hs_prb}; while at $\phi> \phi_{\rm c}$ the bond order parameter $\langle q_6 \rangle$ increases more rapidly due to the emergence of crystalline clusters. Further densification of the HS leads to a strong fluctuations (``oscillations") of the global parameter $\langle q_6 \rangle$, which start at $\phi \simeq 0.68$ and suggest another transition\cite{hs_prb,makse}. This oscillatory behavior is due to the fact that the hcp and the fcc crystalline structures are characterized by quite different values of the bond order parameter $q_6$ (see Table~\ref{t1}); and at a given $\phi$, dense partially crystallized HS packings can be  realized  with quite different abundances of hcp-like and fcc-like spheres.
%Inset (b) in Fig.~\ref{op} shows behavior of the mean $q_6$ in the vicinity of RCP in fine detail;  the dependence $\langle q_6 \rangle(\phi)$ reveals well defined breaking of  the curve slope near the Bernal
%limit (at $\phi \approx 0.6498$). It means, in part, that global $q_6$ can be used as a measure characterizing the RCP transition.
Another important parameter characterizing the local structure is the abundance $N_{\rm ico}$ of spheres having icosahedral-type (fivefold) symmetry (see Fig.~\ref{op} inset). Here, ico-like spheres are defined as spheres having $q_6 \ge 0.61$. The density dependence of $N_{\rm ico}$ shows that ico-like spheres
can exist only in a narrow range of pf, and its abundance maximizes at $\phi_{\rm c}$.

%We note that the appearance of crystalline particles and the disappearance of ico-like particles coincide at the same density.
%correlate nicely (see inset (a) in Fig.~\ref{op} for the comparison) . This suggests that the abundance of ico-like spheres can be used as another measure, characterizing HS in the vicinity of RCP. Seems to be the %ico-like particles prevent crystallization of HS below the RCP limit.

\begin{figure}
\includegraphics[width=12.cm]{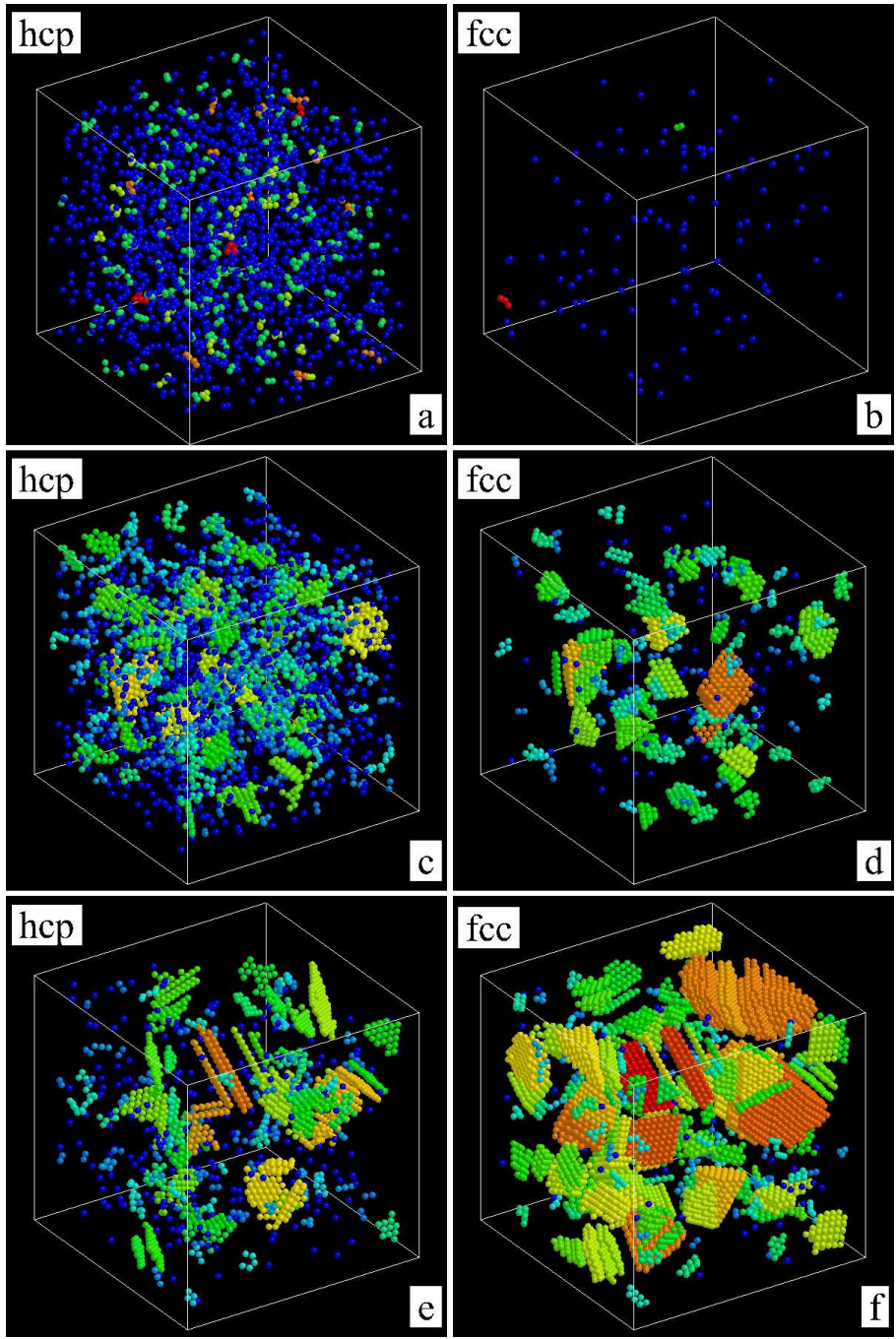}
\caption{(Color online) Distributions of hcp and fcc clusters over the space at different $\phi$ values:  $\phi \simeq 0.65$ (a, b), $\phi \simeq 0.66$ (c, d) and $\phi \simeq 0.68$ (e, f). Particles are color-coded by the mass of the cluster (in the unit of the mass of a single sphere). At $\phi \simeq 0.65$ single hcp-like spheres dominate. At $\phi \simeq 0.65$,  three dimensional local clusters of both hcp and fcc-like particles are observed.  Further densification of 
HS results in a structural transition: the 3D local clusters transform into global and planar layers (e, f).} \label{cl}
\end{figure}

The above analysis provide detailed information about the local and the global orders. Next, we study the abundance of crystalline particles and their spatial arrangement. We use the standard friends-of-friends algorithm\cite{fof} to find crystalline clusters.  Figure~\ref{cl} presents the spatial distribution and the shape of crystalline clusters composed of hcp and fcc-like spheres, at several different $\phi$. When $\phi < 0.65$ (Fig.~\ref{cl}(a, b)), the crystalline particles are mostly isolated; increase of packing fraction $\phi$ leads to an appearance of several 3D (containing tens of spheres, both hcp and fcc-type) clusters (Fig.~\ref{cl}(c, d) at $\phi \approx 0.66$). Upon further densification, these 3D clusters transform into 2D layers spanning system-wide (see Fig.~\ref{cl}(e, f) at $\phi \approx 0.68$) ~\cite{hs_prb}.

\begin{figure}
\includegraphics[width=12.cm]{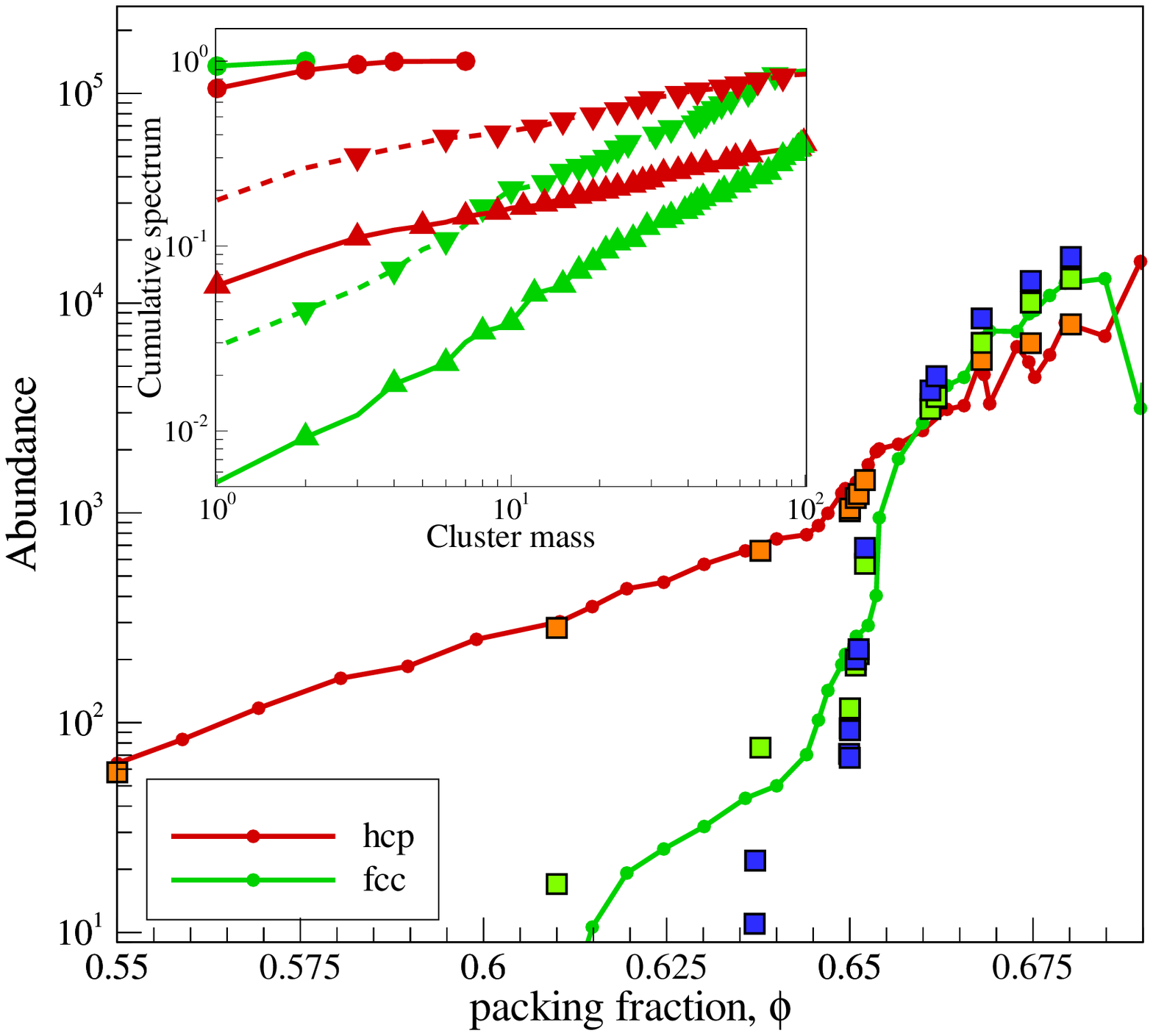}
\caption{(Color online) Abundance of hcp-like (red line-points) and fcc-like (green line-points) spheres versus the packing fraction $\phi$, with $N=10^4$. Results of larger scale simulations with $N =64\times 10^3$ spheres are shown by squares (orange and green symbols correspond to hcp and fcc-like spheres).  Additionally, the abundance of spheres having number of contacts $N_c = 12$ is plotted by blue square. Inset shows the relative cumulative spectrum of hcp (fcc)-like (red and green line, respectively ) clusters versus the cluster mass at different $\phi$ values: $\phi \simeq 0.68$ (solid, triangles), $\phi \simeq 0.66$ (dashed, gradients) and $\phi \simeq 0.65$ (solid, circles).}
\label{cr}
\end{figure}

The abundances of hcp-like ($n_{\rm hcp}$) and fcc-like ($n_{\rm fcc}$) spheres are shown in Figure~\ref{cr} as functions of packing fraction $\phi$. At $\phi \le \phi_c$ the relatively slow increase of $n_{\rm tot} = n_{\rm hcp} + n_{\rm fcc}$ is mainly due to the emergence of randomly distributed single hcp-like spheres. Around $\phi \simeq \phi_c$ the slopes of $n_{\rm hcp}(\phi)$ and $n_{\rm fcc}(\phi)$ increase significantly. At $\phi \simeq 0.66$, the abundance of the fcc-like spheres equals to that of the hcp-like spheres ($n_{\rm fcc} \approx n_{\rm hcp})$. Additionally, the abundance of spheres having number of contacts $N_c = 12$ is plotted in Figure~\ref{cr}, as one more sensitive indicator of the crystallization of a dense HS. The relative cumulative spectrum of cluster mass is plotted in Figure~\ref{cr} inset, showing that this distribution becomes wider as the density increases.

%The hcp/fcc spectrum reveals already mentioned features: only small clusters consisting of few spheres are present in the HS system. As the density increases, larger clusters appear. This observation is consistent with the classical nucleation process\cite{deb}; the denser packings are more ``supercooled'' before falling out of equilibrium, therefore they have lower free energy barriers to form large clusters.

\subsection{Discussion and conclusion}
The above results illustrate a sharp first-order like transition\cite{makse} at $\phi_{\rm c}$ between disordered (frictional) packings and partially ordered packings. Amorphous frictionless packings only exist at a single density $\phi_c$: below $\phi_c$, the packings are unstable unless they are frictional (and the frictional packings are disordered) \cite{makse}; above $\phi_c$, the
packings are partially ordered. On the other hand, multiple amorphous frictionless states have been found in other
simulations\cite{binhs, hdhs}, which supports the mean-field theoretical prediction that amorphous jammed packings should exist in the interval (so-called J-line) $\phi \in [\phi_{\rm th}, \phi_{\rm GCP}]$\cite{rmp1}. We stress that this apparent contrast is due to the presence of crystallization: If the nucleation rate is slow enough, such as in polydisperse\cite{binhs} or large dimensional systems\cite{hdhs}, then crystallization is suppressed and dense amorphous packings can be obtained by slow compressions; while in 3D monodisperse systems, the nucleation process is so fast such that any slow compression would result in partial crystallization. Combined with the mean-field prediction, our results suggest two possible scenarios for 3D monodisperse packings: (i) $\phi_{\rm c} < \phi_{\rm th}$, therefore dense amorphous packings $\phi>\phi_{\rm th}$ are hidden by crystallization; (ii) $\phi_{\rm th} \lesssim \phi_{\rm c}$, however the difference between $\phi_{\rm th}$ and $\phi_{\rm c}$ is very small, such that the amorphous packings in the range $[\phi_{\rm th}, \phi_{\rm c}]$ are difficult to detect based on present numerical accuracy. Further studies are required to test these two scenarios.

To conclude, we have studied structural properties of dense hard spheres in the vicinity of the Bernal limit. The spatial and bound order parameters, and the abundances of hcp, fcc and ico-like particles,  all indicate a sharp disorder-order phase transition at $\phi_{\rm c}$. Finally, we find that at packing fraction $\phi \simeq 0.68$ an additional structural transition occurs: 3D-like shape of solid-like crystalline clusters changes to a planar-like structure.

\begin{acknowledgement}
We thank F. Zamponi and P. Charbonneau for helpful discussions. This study was supported by NSF-CMMT and DOE Geosciences Division. BAK was  supported by  the Russian Science Foundation, Project no. 14-12-01185.
\end{acknowledgement}


\begin{thebibliography}{99}
\bibitem{rmp1} G. Parisi, F. Zamponi, Rev. Mod. Phys, {\bf 82}, 789 (2010).
\bibitem{rmp2} S. Torquato, F.H. Stillinger, Rev. Mod. Phys, {\bf 82}, 2633 (2010).
\bibitem{ncom} A. Baule et al., Nature Comm., {\bf 4}, 2194 (2013).
\bibitem{hs_prb} B. A. Klumov, S. A. Khrapak, G. E. Morfill, Phys. Rev. B {\bf 83}, 184105 (2011).
\bibitem{makse} Y. Jin, H.A. Makse, Physica A,  {\bf 98}, 5362 (2010).
\bibitem{jtb} M. Skoge, A. Donev, F.H. Stillinger, Salvatore Torquato, Phys. Rev. E 74, 041127 (2006).
\bibitem{bernal} J.D. Bernal, Proc. Royal Soc., Series A, {\bf 280}, 299 (1964).
\bibitem{stein} P. Steinhardt, D. Nelson, M. Ronchetti, Phys. Rev. B., {\bf 28}, 784 (1983); 
P. Steinhardt, D. Nelson, M. Ronchetti, Phys. Rev. Lett.,  {\bf 47}, 1297 (1981).
\bibitem{torqa} M.D. Rintoul and S. Torquato, J. Chem. Phys.,  {\bf 105}, 9528 (1996).
\bibitem{tw} P. Wolde, R. Ruiz-Montero, D. Frenkel, J. Chem. Phys.,  {\bf 104}, 9932 (1996).
\bibitem{hs_init} A.S. Clarke, J.D. Wiley, Phys. Rev. B, {\bf 38}, 3659 (1988).
\bibitem{troadec} P. Richard, A. Gervois, L. Oger, J.P. Troadec, EPL,  {\bf 48}, 415 (1999).
\bibitem{q6a} S. Torquato, T.M. Truskett, P.G. Debenedetti, Phys Rev. Lett, {\bf 84}, 2064 (2000);
T.M. Truskett, S. Torquato, P.G. Debenedetti, Phys Rev. E., {\bf 62}, 993 (2000).
\bibitem{iv} I. Volkov at al., Phys. Rev. E, {\bf 66}, 061401 (2002)
\bibitem{torqb} J. R. Errington, P. G. Debenedetti,  Torquato, J. Chem. Phys. {\bf 118}, 2256 (2003).
\bibitem{aste} T. Aste, M. Saadatfar , A. Sakellariou, T.J. Senden, Physica A,  {\bf 339}, 16 (2004).
\bibitem{anik} A.V. Anikeenko, N.N. Medvedev, Phys. Rev. Lett.  {\bf 98}, 235504 (2007).
\bibitem{lj1} B.A. Klumov, JETP Lett. {\bf 97}, 6, 372  (2013).
\bibitem{lj2} B.A. Klumov, JETP Lett. {\bf 98}, 5, 259  (2013).
\bibitem{kapf} S.C. Kapfer, et al., Phys. Rev. E, {\bf 85}, 030301 (2012).
\bibitem{franc} N. Francois et al., Phys. Rev. Lett., {\bf 111}, 148001 (2013).
\bibitem{bar} V. Baranau and  U. Tallarek, Soft Matter, {\bf 10}, 3826  (2014).
\bibitem{nat} M. Rubin-Zuzic {\it et al}., Nature Phys. {\bf 2}, 181 (2006);
B.A. Klumov, {\it et al.}, JETP Lett.,  {\bf 84}, 542 (2006).
\bibitem{3Da} B.A. Klumov, G.E. Morfill, JETP Lett.,  {\bf 96}, 444 (2009);
B.A. Klumov, G.E. Morfill, JETP Lett.,  {\bf 107}, 908 (2008).
\bibitem{mitic} S. Mitic {\it et al}., Phys. Rev. Lett.,  {\bf 101}, 125002 (2008).
\bibitem{3Db} B.A. Klumov {\it et al}., Plasma Phys. Contol. Fusion,  {\bf 51}, 124028 (2009);
B.A. Klumov {\it et al}., EPL,  {\bf 92}, 15003 (2010).
\bibitem{pu} B.A. Klumov, Phys. Usp.,  {\bf 53}, 1053 (2010).
\bibitem{khr} S. A. Khrapak, et al., Phys. Rev. Lett. {\bf 106}, 205001 (2011).
\bibitem{khr1} S. A. Khrapak, et al., Phys. Rev. E {\bf 85}, 066407 (2012).
\bibitem{coll0} U. Gasser, Weeks E.R., Schofield A., Pusey P.N., Weitz D.A., Science,  {\bf 292}, 5515, (2001).
\bibitem{coll1} T. Kawasaki and  H. Tanaka, PNAS, {\bf 107}, 14036 (2010).
\bibitem{pc} O.A. Vasilyev, B.A. Klumov, A.V. Tkachenko, Phys. Rev. E  88 (2013) 012302.
\bibitem{gran} A. Panaitescu, K. Reddy, A. Kudrolli, Phys. Rev. Lett. 108, 108001 (2012).
\bibitem{mg} A. Hirata et al., Science {\bf 341}, 376 (2013).
\bibitem{rss} Y. Fomin et al., J. Chem. Phys. {\bf 141}, 034508 (2014).
\bibitem{t1a} S. Ogata, Phys Rev. A, {\bf 45}, 2, 1122 (1992).
\bibitem{t1b} Y. Wang and C. Dellago, J. Phys. Chem. B, {\bf 107}, 9214 (2003).
\bibitem{t1c} P. Ganesh and M. Widom, Phys Rev. B, {\bf 74}, 134205 (2006).
\bibitem{t1d} K. Subramanian et al., Phys Rev. B, {\bf 74}, 155441 (2006).
\bibitem{Donev2005} A. Donev, S. Torquato, and F. H. Stillinger, Phys. Rev. E {\bf 71}, 011105 (2005).
\bibitem{fof} M. Davis et al., ApJ, {\bf 292}, 371 (1985).
\bibitem{neighbor} Note that in Ref.~\cite{makse}, $N_{\rm nn}$ is equal to the number of neighbors that are in mechanical contact with the central sphere, which is smaller than $N_{\rm nn} =12$ used in this paper. Consequently, the values of order parameters are slightly different.
%This results in a small difference of $\langle q_6 \rangle$ values between Fig.~\ref{op} and Fig. 4(a) in Ref.\cite{makse}.
\bibitem{deb} P. G. Debenedetti, {\it Metastable Liquids}, Princeton University Press, Princeton, (1996).
\bibitem{binhs} P. Chaudhuri, L. Berthier, S. Sastry, Phys. Rev. Lett. {\bf 104}, 165701 (2010).
\bibitem{hdhs} P. Charbonneau, E. Corwin, G. Parisi and F. Zamponi, Phys. Rev. Lett., {\bf 109}, 205501 (2012).
{\it et al}, Phys. Rev. E {\bf 60}, 4551 (1999); T. Aste, J. Phys. Cond. Matter,  {\bf 17}, S2631 (2005).
\end{thebibliography}
\end{document}